\documentstyle[aps,prl,epsf,twocolumn]{revtex}
\begin{document}
\title{Anomalous magnetic response of the valence-fluctuating compound
YbInCu$_4$}
\author{J. K. Freericks and V. Zlati\'c$^{\dag}$}
\address{Department of Physics, Georgetown University, Washington, DC 20057}
\date{\today }
\maketitle

\begin{abstract}
The exact solution of the spin one-half Falicov-Kimball model, with random 
hopping between the lattice sites, is used to explain the anomalous magnetic 
response of Yb-based valence-fluctuating intermetallic compounds. The
anomalous behavior arises from an entropy-driven local-moment--nonmagnetic
transition of unhybridized Yb ions in these materials which can also be used to
explain the observed metamagnetism and resistivity anomalies.
\end{abstract}

\renewcommand{\thefootnote}{\copyright}
\footnotetext{ 1996 by the authors.  Reproduction of this article by any means
is permitted for non-commercial purposes.}
\renewcommand{\thefootnote}{\alpha{footnote}}


The valence-fluctuating (VF) compounds YbInCu$_4$, YbIn$_{1-x}$Ag$_x$Cu$_4$ 
and Yb$_{1-x}$In$_x$Cu$_2$ exhibit large anomalies in their
thermodynamic \cite{felner.86,yoshimura.88,kindler.94}, 
spectroscopic \cite{felner.86,kojima.90,nakamura.90} 
and transport properties \cite{felner.86,kindler.94,sarrao.96},
around a characteristic, sample-dependent temperature $T_V$ ($30-70 K$). 
The anomalies are due to the VF, as the $L_{III}$-edge, 
M\"ossbauer, X-ray absorption  and the thermal-expansion data 
show that the Yb ions fluctuate between a $2+$ and $3+$ state. 
However, the atomic volume is altered at $T_V$ by a small amount and 
the average valence of the Yb ions\cite{felner.86,kindler.94,nakamura.90} 
does not change by more than 4\%.
The X-ray and neutron data do not show any structural changes across the VF 
transition\cite{kindler.94,kojima.90,sarrao.96,lawrence.96},  
and the neutron\cite{kojima.90} and the NMR data\cite{yoshimura.88} 
rule out magnetic ordering as a possible 
explanation of the anomalous behavior.
Here, we focus on the anomalous magnetic response of Yb-based VF 
compounds, as revealed by the temperature dependence of the 
low-field susceptibility, $\chi(T)$, 
and the field-induced torque, $\Gamma(T)$, 
and by the measurements of the high-field magnetization, $M(H)$, 
at temperatures below $T_V$. 

The anomaly in $\chi(T)$ is seen most clearly 
in the YbInCu$_4$ single-crystal data \cite{kindler.94}, 
as a pronounced asymmetric peak just above $T_V$. 
The anomalous peak rises from a large uniform background and is much steeper 
on the low-temperature than on the high-temperature side. 
Below $T_V$, $\chi(T)$ drops to a minimum and then 
grows slowly towards the lowest temperatures, 
where a weaker Curie-like upturn is often seen. 
For $T\gg T_V$, $\chi(T)$ decays in a 
Curie-Weiss fashion but the magnitude of the susceptibility 
at $300\;K$ is about the same as at the low-temperature minimum. 
Surprisingly, the magnitude, and even the sign, of the Curie-Weiss 
temperature appears to be sample dependent. 
Such behavior is quite different from what one finds in a typical metallic
heavy-fermion sample, like YbAgCu$_4$ \cite{rossel.1987,yoshimura.88}, 
where the low-temperature response is much enhanced over the room-temperature 
one and $\chi(T)$ exhibits a shallow and symmetric maximum.

The torque induced by the field
(measured for the single crystals used in Ref.\cite{kindler.94}) 
shows that the magnetic susceptibility is also anisotropic\cite{miljak.96}, 
which is quite unusual for compounds of cubic symmetry.
The torque, $\Gamma(T)$,
is characterized by an asymmetric peak, which is clearly related to 
the anomalous peak in $\chi(T)$. Below $T_V$, $\Gamma(T)$ drops by 
several orders of magnitude and then, like $\chi(T)$, it shows an upturn 
as the temperature is further reduced. The decay of $\Gamma(T)$ above $T_V$ is 
very rapid and cannot be described by a Curie-Weiss law. 
The $\Gamma$-anomaly, like the $\chi$-anomaly, is 
sharper and more asymmetric for the samples with lower $T_V$'s. 

The anomaly in the high-field  magnetization of 
YbInCu$_4$ and YbIn$_{1-x}$Ag$_x$Cu$_4$ 
\cite{yoshimura.88,katori.94} is characterized by a sudden increase 
of the slope and the saturation of $M(H)$ at about $H_V\simeq 30 - 50\;T $. 
The magnetostriction data\cite{yoshimura.88} indicate that 
the metamagnetism relates to valence fluctuations of Yb ions  
but the field-induced change in the average 
f-count, $\left<n_f\right>$ is even smaller than the temperature-induced one. 
At low temperatures, the Zeeman energy at $H_V$ is comparable to the 
thermal energy at $T_V$. At elevated temperatures, the metamagnetic 
transition is absent\cite{kojima.90}. 
The magnetic anomalies at $T_V$ and $H_V$, 
and the characteristic energies $k_B T_V$ and $\mu_B H_V$ are sample 
dependent\cite{yoshimura.88}, while the low-field response, for $T < T_V$, 
is quite robust with respect to doping and annealing.

The simultaneous analysis of the susceptibility and the torque data 
shows\cite{miljak.96} that the magnetic response is due to 
two physically distinct components. 
The first one, $\chi_{VF}$, is large (on an absolute scale), isotropic, 
and sample independent. 
The second one, $\chi_{local}$, is vanishingly small below $T_V$ and 
smaller than $\chi_{VF}$ at high temperatures,
but dominates $\chi(T)$ and $\Gamma(T)$ near $T_V$.  
$\chi_{local}$ is anisotropic and strongly sample dependent. 
In addition, the field and sample dependence of the $M(H)$ 
data also supports the picture of two types of magnetic excitations. 

To explain these features we recall that the Yb compounds mentioned above 
crystallize in $C15b$-(AuBe$_5)$-type structure\cite{kojima.90,lawrence.96} 
with In and Yb ions at the so-called 4a and 4c sites, respectively. 
Because of the short 4a-4c distance, the f-states of the Yb ions
hybridize with the In 
ions and give rise to an enhanced background susceptibility, $\chi_{VF}$, 
which dominates $\chi(T)$ much above and below $T_V$. 
For $T\leq T_V$, these hybridized states also yield the sample-independent 
low-field magnetization. 
On the other hand, the disordered
C15-type structure, in which the 4a and 4c sites 
are randomly occupied by Yb and In ions, is quite close to the 
C15b-type structure, and is not ruled out by 
the X-ray and neutron data \cite{kojima.90,lawrence.96} (especially for 
a small number of disordered Yb ions occupying the In 4a sites).  
Thus, we assume that a small number $(N)$ of Yb ions in YbInCu$_4$ 
occupy the In 4a sites {\it and that their 4f-states remain unhybridized}. 
The hybridization between these ill-placed 4a-Yb ions and the regular 
4a-In ions is absent because the 4a-4a distance is nearly twice the 
4a-4b distance. 
The magnetization, of the 4a-Yb ions in the $f^{13}$ configuration 
is angle-dependent, because the $g$-factor of the unhybridized $f$-holes 
becomes anisotropic in the distorted crystal environment. 
The external field induces the torque, which relates to the 
local susceptibility as, 
$\Gamma\simeq 
\frac{\Delta g^2}{\left<g^2\right>}\left<\chi_{local}(N,T)\right>H^2$, 
where $\Delta g^2=g^2_x - g^2_y$ is the $g$-factor anisotropy 
between the x and y direction, and $\left<\cdots\right>$ 
denotes the angular average. 
As show below, the unhybridized Yb ions crossover at $T_V$  
from the magnetic $3+$ to the nonmagnetic $2+$ configuration, so that  
$\chi_{local}$ and $\Gamma$ vanish. At low temperatures and high fields 
the unhybridized states lead to the metamagnetic transition with a  
small energy difference between the low-field and the high field states. 
Note, $\left<n_f\right>$ does not change much either at $H_V$ or at $T_V$, 
because the VF transition involves only $N$ 4a-impurities out of a total of
$N_l$ Yb ions, i.e. $N/N_l\ll 1$. 

Our model for the anomalous magnetic response of these materials begins
with this assumption that the f-electrons of the disordered Yb ions 
are {\it not} hybridized with the rest of the lattice.  These 
ions form a random lattice.  The conduction electrons
can move, via nearest-neighbor hopping on the regular
lattice, between any of these random lattice sites; 
i.e. the system can be modeled by localized f-states at energy $E$ 
and delocalized d-states, with a {\it random} hopping between each of
the sites of the lattice. 
We assume  that the $f$ and $d$ states have a common chemical potential 
($\mu$) and interact by a Coulomb repulsion ($U$). In addition, 
both the $d$ and $f$ particles carry a spin label $\sigma$ and the 
$d$-level can accommodate $2$ electrons (or holes) of the opposite spin.
The occupancy of the $f$-level is restricted to $n_f\leq 1$
because of the large Coulomb repulsion ($U_{ff}\sim\infty$) of the 
$f$-particles of opposite spins. 
To discuss the Yb-based VF compounds we use the hole-picture, in 
which $E < \mu$ and the total number of holes at the ill-placed sites 
is restricted to $n_h=n_d^h + n_f^h \leq 3$. 
In the electron-picture, one has $E > \mu$ and restricts the total 
number of electrons  to $n_e=n_d^e + n_f^e \leq 3$. 
The magnetic field $h$ couples to the $f$ and $d$ states 
but with different $g$-factors ($g=4.5$ for the $f$-holes). 
This picture is described by a spin one-half Falicov-Kimball (FK)
model \cite{falicov.69}, 
\begin{eqnarray}
\label{eq:H_FK}
H&=&\sum_{ij,\sigma} (t_{ij} - \mu \delta_{ij})
d_{i\sigma}^{\dagger} d_{j\sigma} + \sum_{i,\sigma} (E-\mu)
f_{i\sigma}^{\dagger} f_{i\sigma} \cr
&+& U \sum_{i,\sigma \sigma^{\prime}} 
d_{i\sigma}^{\dagger} d_{i\sigma} 
f_{i\sigma^{\prime}}^{\dagger} f_{i\sigma^{\prime}}  + U_{ff} \sum_{i,\sigma} 
f_{i\uparrow}^{\dagger} f_{i\uparrow} f_{i\downarrow}^{\dagger} f_{i\downarrow}
\cr
&-&\mu_Bh\sum_{i,\sigma}\sigma (2d_{i\sigma}^{\dag}d_{i\sigma}+
gf_{i\sigma}^{\dag}f_{i\sigma})\quad ,
\end{eqnarray}
where the notation of Ref.\cite{brandt.1989} is used and $i$ and $j$ 
denote the sublattice sites where the ill-placed Yb ions are located.  
The hopping matrix elements $t_{ij}$ are chosen from a random distribution
in such a fashion that as $N\rightarrow\infty$ the noninteracting
density of states becomes Gaussian $\rho(\epsilon)=t^*\exp[-\epsilon^2/t^{*2}]/
\sqrt{\pi}$ \cite{dos}. The only role of the regular lattice, with  
$N_l$ sites, is to provide an infinite-range (random) connectivity between 
the sublattice sites.

In the limit $N\rightarrow\infty$ \cite{kotliar}, and $N/N_l \ll 1$, 
our choice for $t_{ij}$ maps this problem onto the infinite-dimensional 
(local) one, which allows the magnetic susceptibility to be evaluated 
exactly using the methods developed by Brandt and Mielsch \cite{brandt.1989}. 
The only differences we have is that we fix the
total electron concentration (by adjusting $\mu$) rather than fixing
just the $f$-electron concentration.  
The Green's functions and susceptibilities are determined numerically 
by solving the self-consistent equations of Ref.~\cite{brandt.1989}. 
The properties of the model depend on $n_h$, $E-\mu$, and $U$; 
the value of the effective matrix element for random hopping ($t^*$) 
defines the energy scale.  The uniform f-f spin-spin susceptibility
satisfies $\chi_{SDW}^{ff}(T)=n_f^h(T)/2T$. 

We present the susceptibility results of our numerical solutions for 
a variety of different cases in Figure 1.  
Each figure corresponds to a different value of $n_h$, 
with ten values of $E/t^*$ ranging from 0 to $-4.5$ in steps of $-0.5$. 
Fig.~1(a) is a typical high-density result. 
There are 2.5 holes per impurity site, implying $n_f^h\neq 0$ 
at all temperatures, leading to a Curie 
contribution to the susceptibility at small $T$.  
In this regime, the results are rather insensitive to $U$ ($U=10t^*$ here). 
Fig.~1(b) is the $n_h=2.1$ case.
We chose $U=t^*$ here, and the results depend sensitively on $E$, showing a 
downturn at moderate $T$, before the Curie-law divergence sets in for  
$T\rightarrow 0$. 
As $n_h$ is decreased to 2 and beyond, it is no longer necessary for there
to be any $f$ holes remaining at $T=0$, and $\chi^{ff}$ can vanish in that 
limit. 
In Fig.~1(c) we plot $\chi^{ff}$ for $n_h=2$ and $U=2t^*$, 
showing an asymmetric peak that increases in size and sharpens 
as $T_V$ decreases, and which still has a weaker Curie upturn at the lowest 
temperatures, because $n_f^h\ne 0$ as $T\rightarrow 0$.
Here, the results depend quite sensitively on $E$ and $U$, with
the peak lowering in magnitude and broadening as $U$ increases,
and the low-temperature upturn becoming more prominent.  
The case $n_h=1.9$ is shown in Fig.~1(d) for $U=10t^*$.  
Once again we have a sensitive dependence on the parameters, 
but the low-temperature upturn has disappeared.
In some cases $\chi^{ff}$ has a double-hump structure.  
Finally the low-density regime is plotted in Fig.~1(e) 
($n_h=0.5$, $U=10t^*$). Here, as in the high-density limit, 
the results are insensitive to $U$. 

The model also
exhibits a metamagnetic transition, for large enough magnetic fields. 
This is shown in Figure 2 for one case: 
$n_h=2$, $E=-t^*$, $U=2t^*$, and eight temperatures 
from $0.05t^*$ up to $6.4t^*$.   The numerical analysis shows that 
at $T=0$, the Zeeman energy at the transition is of the order of $k_BT_V$. 
$H_V$ decreases at higher temperatures but the the transition 
becomes smoother.
The metamagnetism disappears as $T$ is increased beyond $T_V$, because
the $f$-holes are already occupied in the zero-field limit.

The exact solution reproduces well the overall behavior of the 
experimental data. Figs.(1c)-(1d) capture most of the features 
shown by $\chi(T)$ in 
Refs.\cite{felner.86,yoshimura.88,kindler.94,sarrao.96,lawrence.96,miljak.96}, 
such as an asymmetric peak at $T_V$ that increases in magnitude and sharpens
as $T_V\rightarrow 0$,
while Fig.(2) explains $M(T)$ reported in \cite{yoshimura.88,katori.94},
with a metamagnetic transition that smoothes out and then disappears as
$T$ is increased. 
Various samples will have different numbers of  
Yb impurities and will require different values of the coupling constants. 
Our analysis shows that the lower the transition temperature, 
the more pronounced and steeper the anomaly. 
For large fields, unhybridized Yb ions crossover to a magnetic configuration 
at much lower temperatures, than in the absence of the field. 

Although this is a complicated many-body system, 
in which the behavior depends strongly on the total number of particles 
at the impurity sites and the double occupancy of $4f$-states is prohibited,  
the anomalous behavior can be understood in terms 
of simple thermodynamic considerations. 
For $E<\mu$ and $n_h < 2$, the nonmagnetic state of an Yb impurity 
is energetically more favorable than the magnetic state, so the 
ground state corresponds to the $4f^{14}$ configuration (no $f$-holes). 
However, at high $T$, the large entropy of the $f$-hole spins 
favors the  magnetic state. 
Thus, the high-temperature response of an ensemble of Yb impurities 
approaches the limit $\chi_{local}(N,T)\simeq N\chi_{0}(T)$, 
where $\chi_{0}(T) \simeq (g\mu_B)^2/T$ is the local susceptibility 
of a single Yb ion with one unhybridized magnetic hole.  
The $g$-factor anisotropy of that hole 
gives rise to the field-induced torque which depends on temperature 
via $\chi_{local}(T)$. The anomalous part of the susceptibility and the 
torque grow rapidly as the temperature is reduced until, close to $T_V$, 
the entropy gain is insufficient to compensate the 
energy loss and the $f$ holes are filled to form a nonmagnetic state.  
Thus, $\chi_{local}$ drops significantly, which is the origin of the
sharp asymmetric peak in the $\Gamma(T)$ and $\chi(T)$ data. 
This anomalous peak conceals the broad maximum which is usually found in the 
susceptibility of VF compounds (from the electrons on the regular lattice). 
For $n_h\geq 2$, however, the $f$-hole cannot be completely filled at $T=0$ 
and the residual $f$-spin leads to a low-temperature upturn of $\chi(T)$, 
which is also often seen in the data
\cite{felner.86,yoshimura.88,kindler.94,sarrao.96,lawrence.96,miljak.96}
(this theoretically predicted Curie tail will add to the Curie tail present
in all samples due to impurities).    
The associated Curie constant is very small for $n_h\simeq 2$ but   
increases rapidly as the density of holes increases. 
In YbInCu$_4$, the proximity of the f-level to $\mu$ gives rise 
both to the FK-transition of unhybridized Yb ions and the VF behavior 
of hybridized Yb ions. Had $E-\mu$ been much larger than $t^*$, then
the unhybridized f-levels would not change from $3+$ to $2+$ at low
temperatures.

The transport properties are also explained by this scheme:
the Kondo scattering for $T>T_V\gg T_K$ is at the spin-disorder limit, 
while for $T<T_V$ the free spins vanish and the Kondo scattering is absent. 
Hence, the resistivity jumps at $T_V$ from its small low-temperature 
value to a high-temperature spin-disordered value \cite{kindler.94}.   
The switching on and off of the Kondo scattering
also leads to large anomalies in the elastic constants
through the electron-phonon coupling at the regular lattice sites.  

In summary, the response of YbInCu$_4$ and other Yb-based VF compounds
is analyzed in terms of two independent magnetic excitations.  
The magnetic anomalies are attributed to an entropy-driven transition 
of disordered Yb and explained by the Falicov-Kimball model 
with random hopping. Although the disordered Yb ions undergo 
at $T_V$ a large valence change, $\left< n_f\right>$ does not change much 
since  $N/N_l\ll 1$. Thus we account for 
the correlation between $\chi(T)$ and $\Gamma(T)$,  
explain the metamagnetic transition, and reconcile the large changes 
in magnetic, transport, and elastic properties 
with small changes in the $f$-count and the average atomic volume. 

We acknowledge useful conversations with 
I. Aviani, Z. Fisk, M. Jarrell, B. L\"uthi, M. Miljak, and J. Sarrao. 
J.~K.~F. acknowledges the Donors of The Petroleum Research Fund, administered 
by the American Chemical Society, for partial support of this research 
(ACS-PRF No. 29623-GB6) and the Office of Naval Research Young Investigator
Program (N000149610828).
This project has been funded in part by the National Research Council
under the Collaboration in Basic Science and Engineering Program. The
contents of this publication do not necessarily reflect the views or
policies of the NRC, nor does mention of trade names, commercial
products, or organizations imply endorsement by the NRC.

$^{\dag}$ Permanent address: Institute of Physics of the University of
Zagreb, Croatia.

\begin{figure}[t]
\caption{Magnetic susceptibility of the $f$-holes in the Falicov-Kimball
model. The different curves correspond to $E/t^*=0.0, -0.5,\ldots,-4.5$ (in
general $E$ decreases from top to bottom in these figures).
The different figures are: (a) $n^h=2.5$ and $U=10t^*$; (b) $n^h=2.1$ and 
$U=t^*$;
(c) $n^h=2.0$ and $U=2t^*$; (d) $n^h=1.9$ and $U=10t^*$; and (e) $n^h=0.5$ and 
$U=10t^*$.}
\end{figure}

\begin{figure}[t]
\caption{Total magnetization of the Falicov-Kimball model as a function
of magnetic field.  The different curves correspond to different temperatures:
$T/t^*=0.05,0.1,0.2,\ldots,3.2,6.4$ (the temperature increases from top to 
bottom
in the large-$h$ range).  The parameters are $n^h=2$, $E=-t^*$, and $U=2t^*$.}
\end{figure}

\begin{figure} 
\epsfxsize=3.0in
\epsffile{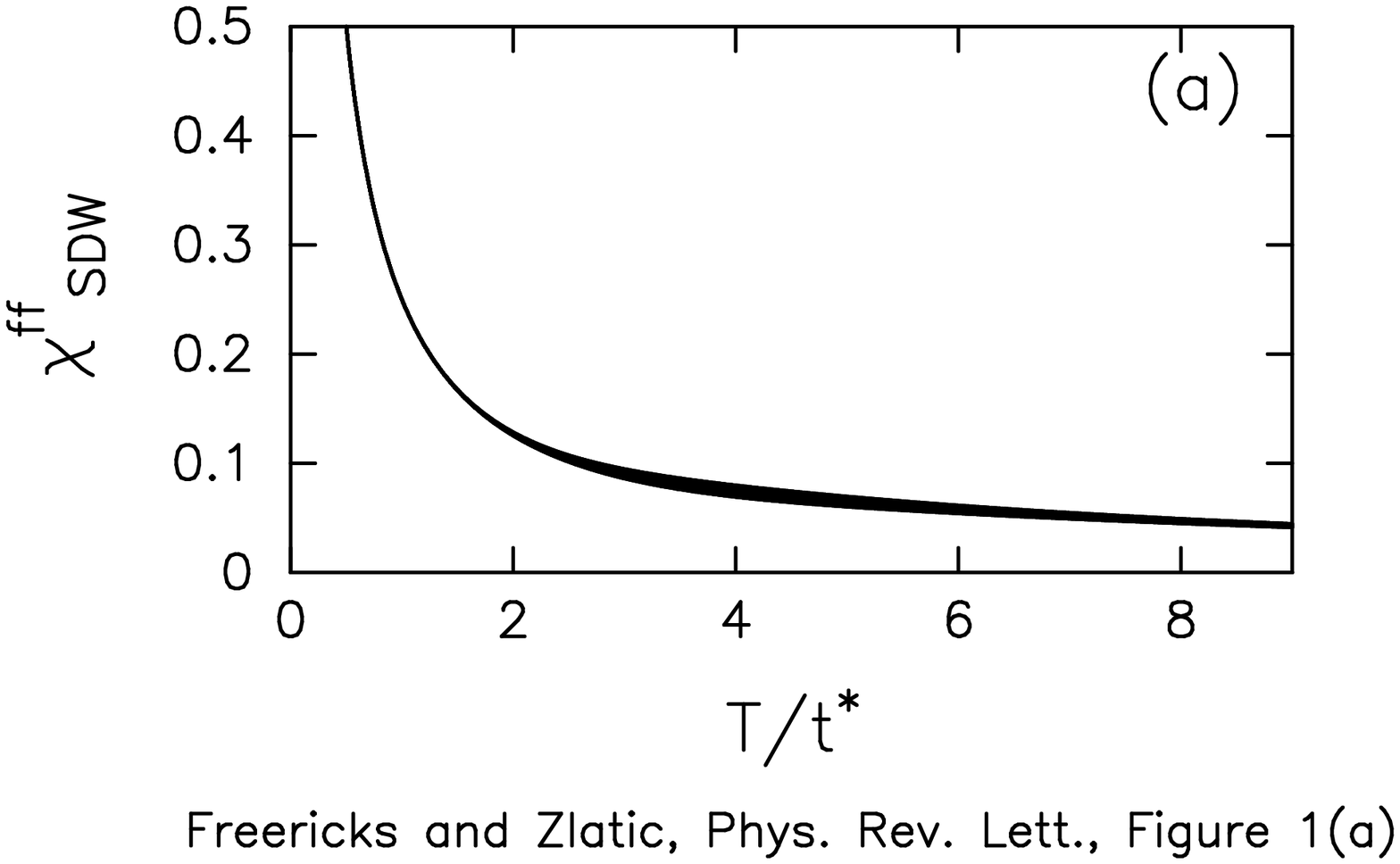}
\end{figure}

\begin{figure} 
\epsfxsize=3.0in
\epsffile{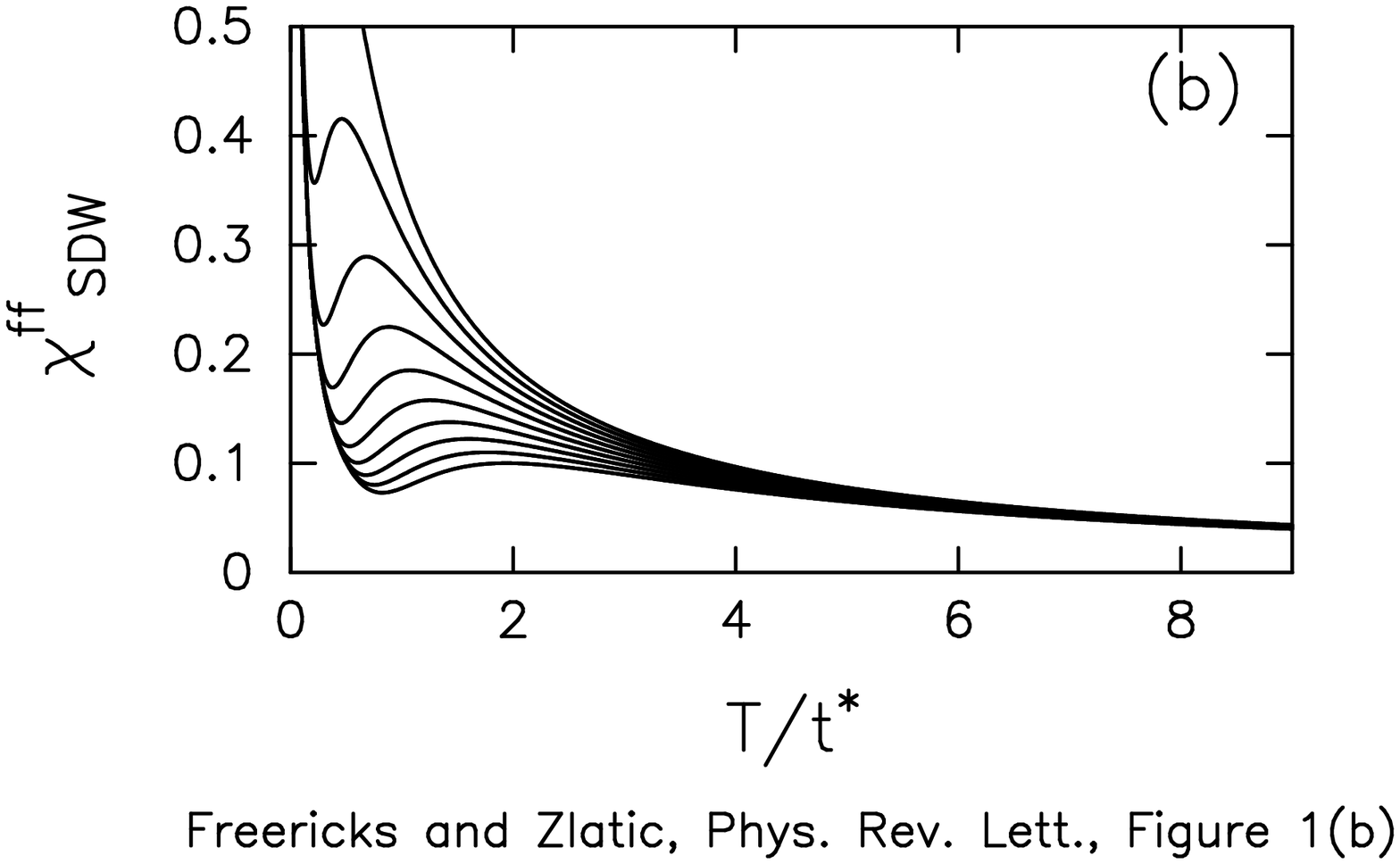}
\end{figure}

\begin{figure} 
\epsfxsize=3.0in
\epsffile{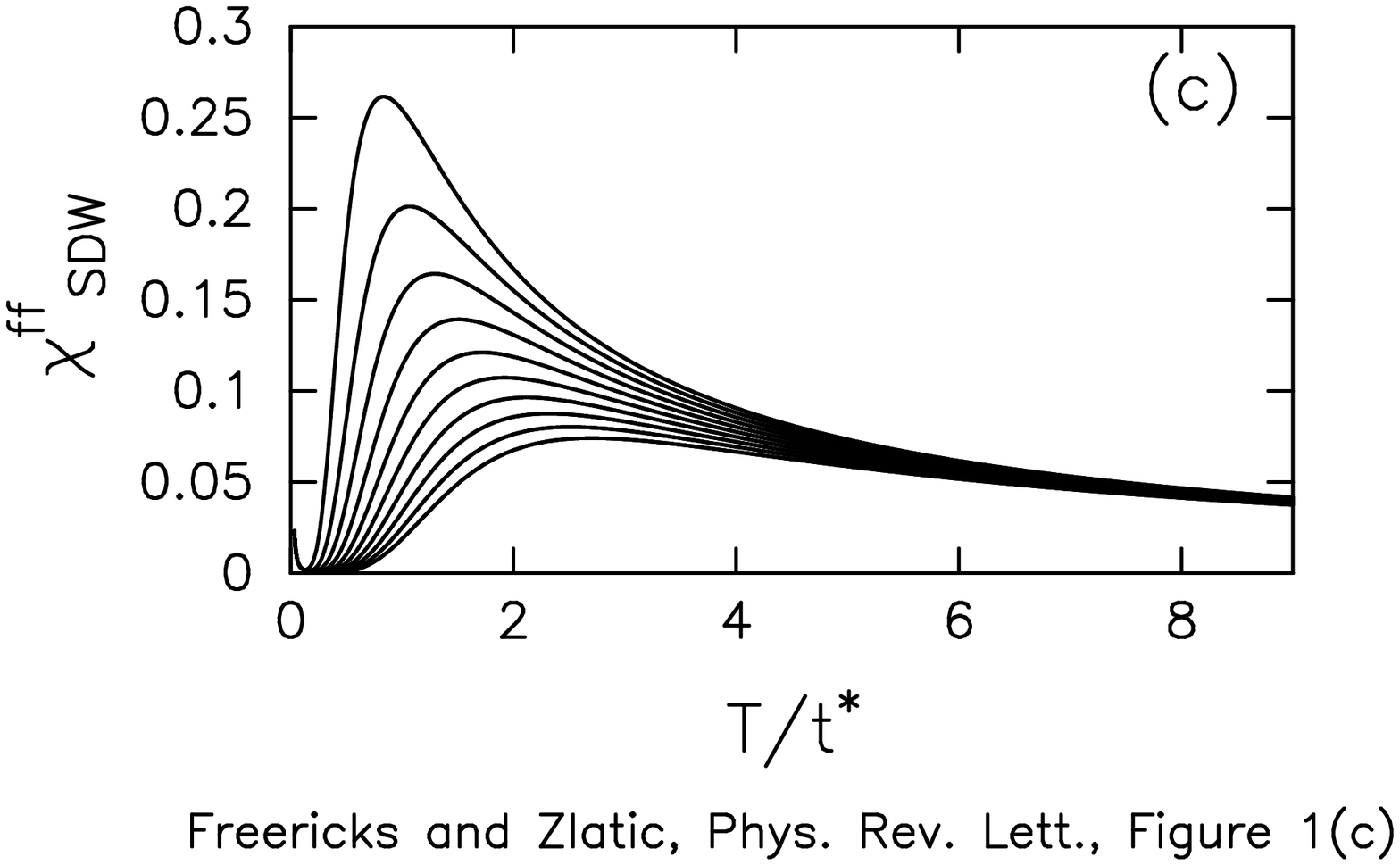}
\end{figure}

\begin{figure} 
\epsfxsize=3.0in
\epsffile{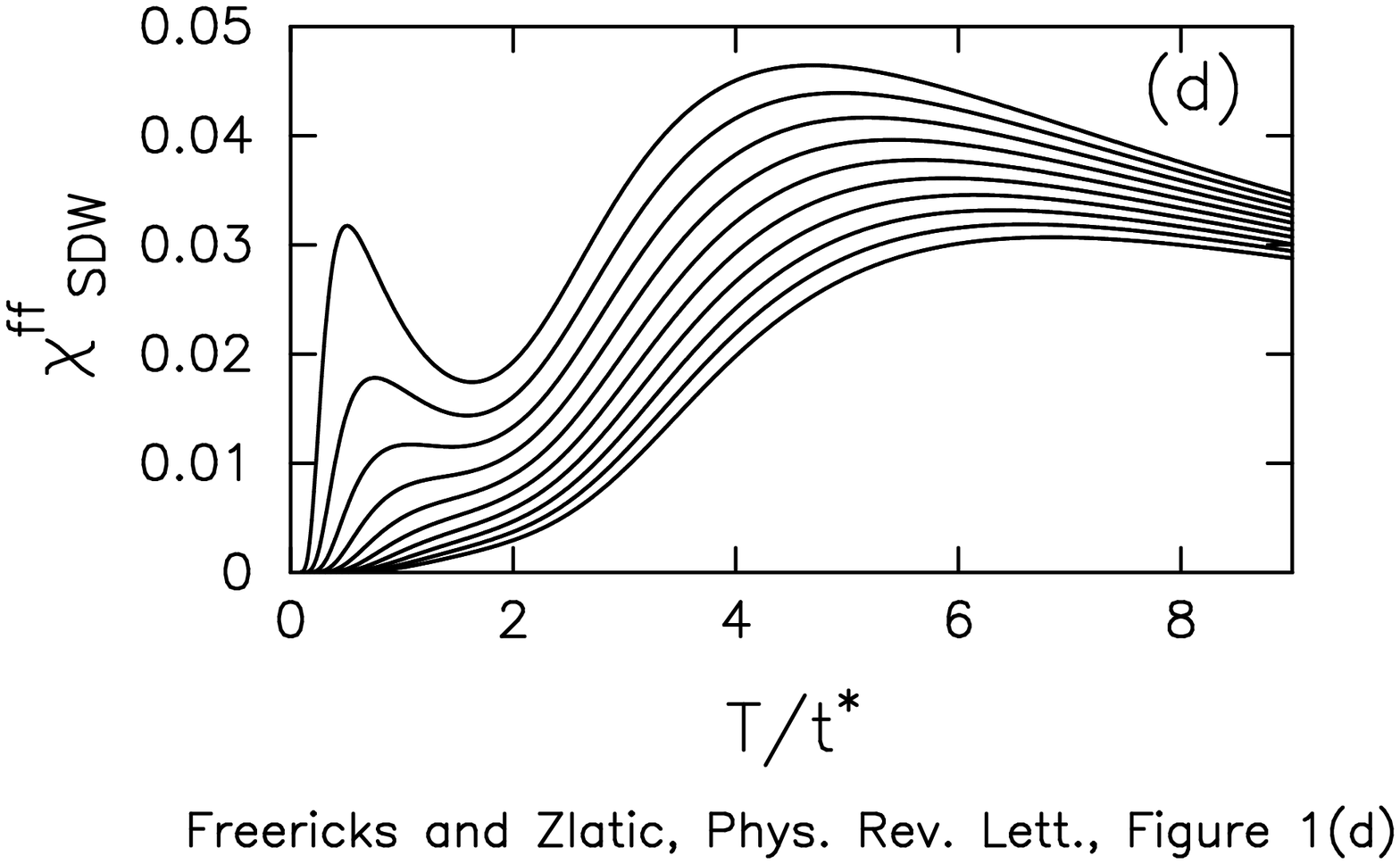}
\end{figure}

\begin{figure} 
\epsfxsize=3.0in
\epsffile{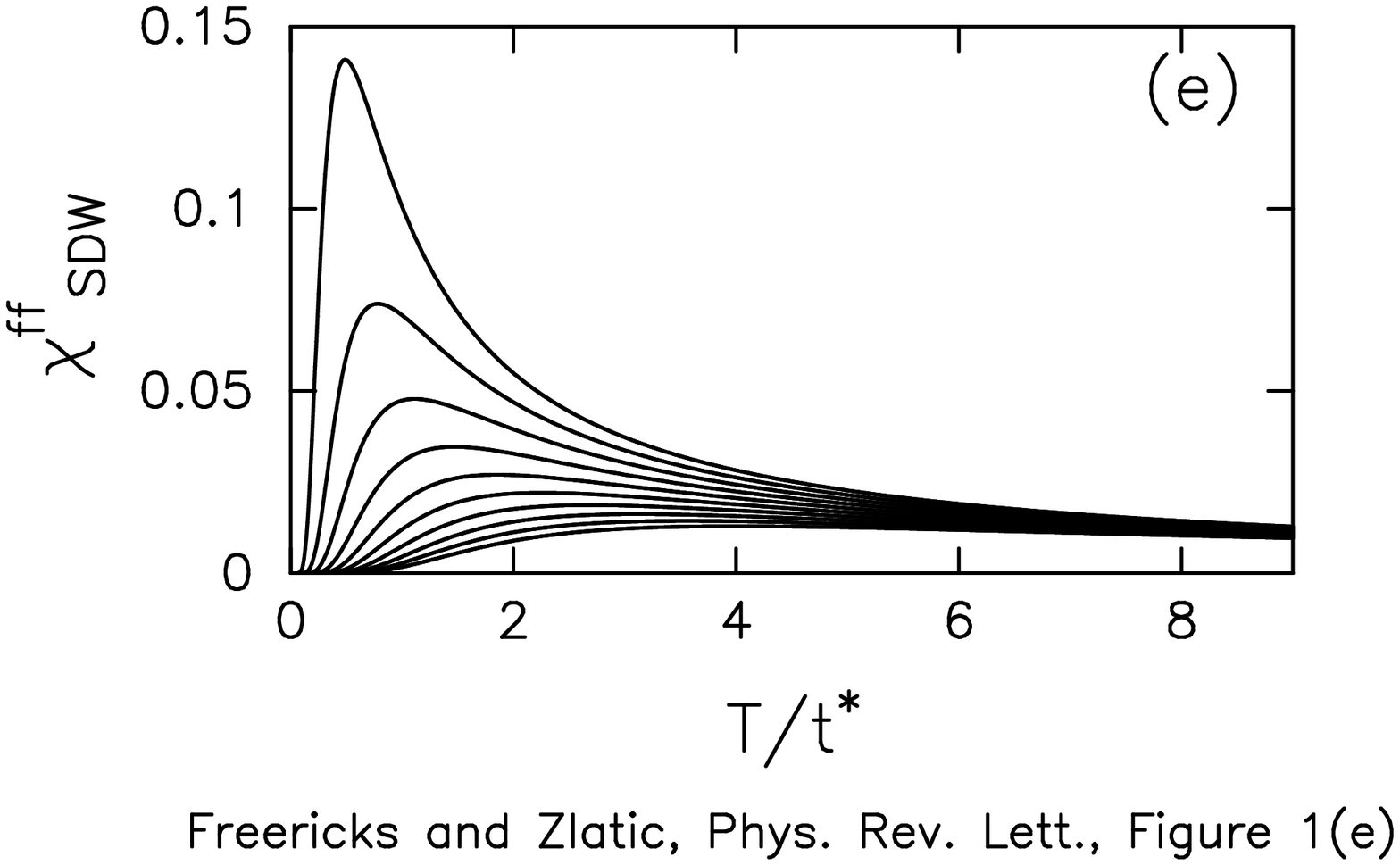}
\end{figure}

\begin{figure} 
\epsfxsize=3.0in
\epsffile{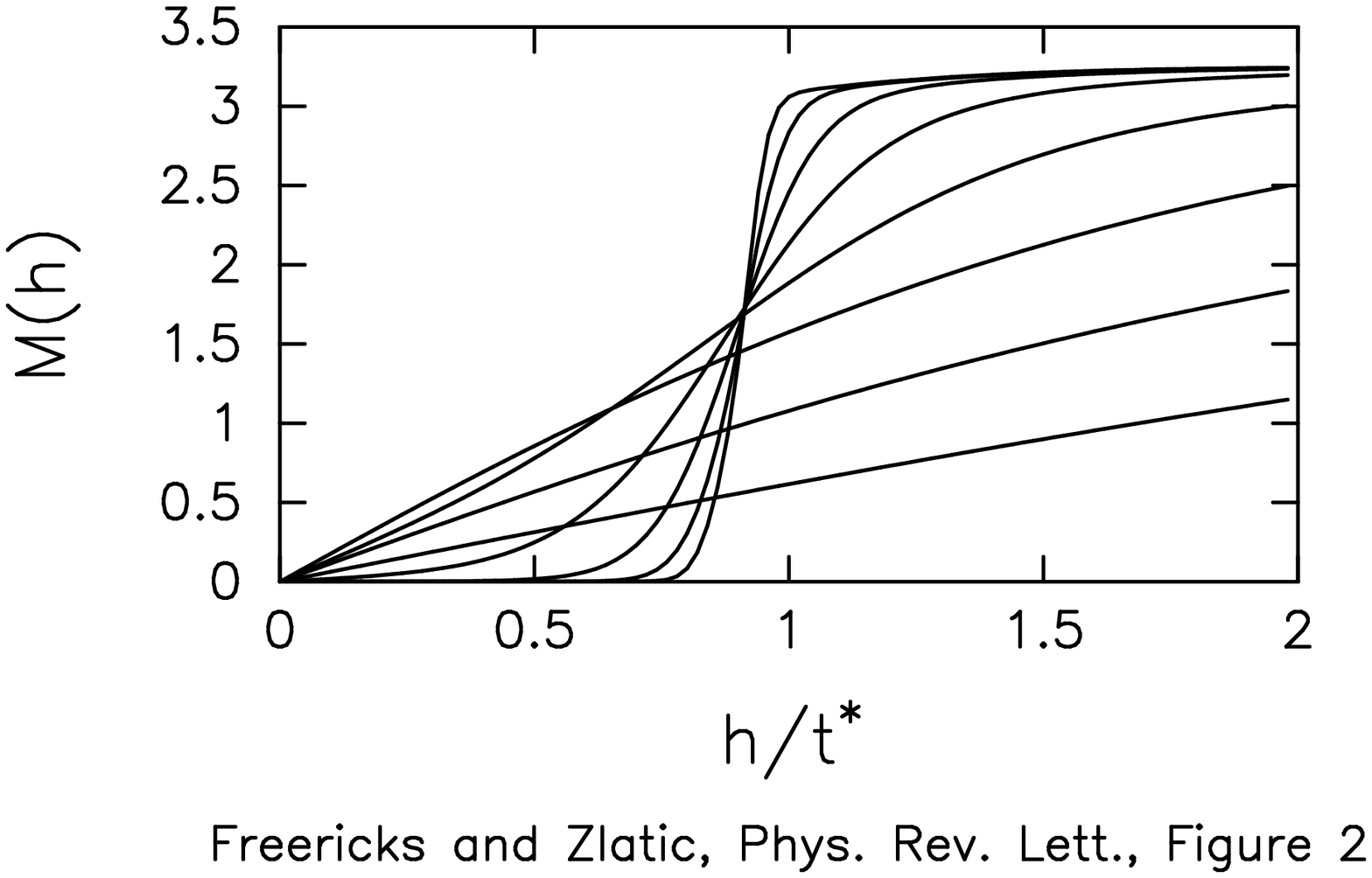}
\end{figure}

\end{document}